\UseRawInputEncoding
\documentclass[%
reprint,
groupedaddress,
showpacs,preprintnumbers,
 amsmath,amssymb,
 aps,
pra,
]{revtex4-2}

\usepackage{graphicx}
\usepackage{epstopdf}
\usepackage{dcolumn}
\usepackage{bm}
\usepackage{gensymb}
\usepackage{textcomp}
\usepackage{fontenc}
\usepackage{upgreek}
\usepackage{float} 
\usepackage{xspace} 
\usepackage[%
	breaklinks=true,
	colorlinks=true,
	linkcolor=blue,
	urlcolor=blue,
	citecolor=blue,
	]{hyperref}
\usepackage{color}
\usepackage{subfigure}

\begin{document}
	
\title{Double Ionization to CO$_2$ Produces Molecular Oxygen: A Roaming Mechanism}
		
\author{Qibo Ma}
\altaffiliation{These authors contributed equally to this work.}
\affiliation{MOE Key Laboratory for Nonequilibrium Synthesis and Modulation of Condensed Matter, School of Physics, Xi'an Jiaotong University, Xi'an 710049, China}

\author{Xintai Hao}
\altaffiliation{These authors contributed equally to this work.}
\affiliation{MOE Key Laboratory for Nonequilibrium Synthesis and Modulation of Condensed Matter, School of Physics, Xi'an Jiaotong University, Xi'an 710049, China}
	
\author{Jiaqi Zhou}
\email[]{zhoujiaqi@xjtu.edu.cn}
\affiliation{MOE Key Laboratory for Nonequilibrium Synthesis and Modulation of Condensed Matter, School of Physics, Xi'an Jiaotong University, Xi'an 710049, China}
	
\author{Xiaorui Xue}
\affiliation{MOE Key Laboratory for Nonequilibrium Synthesis and Modulation of Condensed Matter, School of Physics, Xi'an Jiaotong University, Xi'an 710049, China}

\author{Qingrui Zeng}
\affiliation{MOE Key Laboratory for Nonequilibrium Synthesis and Modulation of Condensed Matter, School of Physics, Xi'an Jiaotong University, Xi'an 710049, China}
	
\author{Peng Li}
\affiliation{MOE Key Laboratory for Nonequilibrium Synthesis and Modulation of Condensed Matter, School of Physics, Xi'an Jiaotong University, Xi'an 710049, China}
	
\author{Lei Wang}
\affiliation{MOE Key Laboratory for Nonequilibrium Synthesis and Modulation of Condensed Matter, School of Physics, Xi'an Jiaotong University, Xi'an 710049, China}
	
\author{Xueguang Ren}
\email[]{renxueguang@xjtu.edu.cn}
\affiliation{MOE Key Laboratory for Nonequilibrium Synthesis and Modulation of Condensed Matter, School of Physics, Xi'an Jiaotong University, Xi'an 710049, China}

	
\date{\today}
	
\begin{abstract}
We report a combined experimental and theoretical study on the formation of O$_2$$^+$ by electron-impact double ionization and fragmentation of carbon dioxide (CO$_2$) molecule. Through fragment ions and electron coincidence momentum imaging, we determine unambiguously the ionization mechanism by measuring the projectile energy loss in association with the C$^+$ + O$_2$$^+$ channel. Further potential energy and trajectory calculations enable us to elucidate the dynamical details of this fragmentation process, in which a bond rearrangement pathway is found to proceed via the structural deformation to a triangular intermediate. Moreover, we demonstrate a new roaming pathway for formation of O$_2$$^+$ from CO$_2$$^{2+}$, in which a frustrated C-O bond cleavage leaves the O atom without sufficient energy to escape. The O atom then wanders around varied configuration spaces of the flat potential energy regions and forms a C-O-O$^{2+}$ intermediate prior to the final products C$^+$ + O$_2$$^+$. Considering the large quantities of free electrons in interstellar space, the processes revealed here are expected to be significant and should be incorporated into atmospheric evolution models. 



\end{abstract}
	
\maketitle
	
Molecular oxygen (O$_2$), as a crucial component of the Earth's atmosphere, is the cornerstone of life-sustaining processes. Prior to the Great Oxygenation Event caused by photosynthetic oxygen production ~\cite{Lyons2014,Bekker2004, james1993,Canfield2005}, abiotic processes dominated Earth's primordial oxygen budget, providing O$_2$ required for the earliest chemical evolution of life. Over the past decades, the combination of two O atoms via three-body collisions (O + O + M $\rightarrow$ O$_2$ + M) has been widely recognized as the main abiotic pathway for O$_2$ production~\cite{Harman2015,Kasting1984,Kasting1981}, in which a third body (M) is involved to carry away the excess energy. The O atoms are generated through the dissociation of CO$_2$ or other precursor molecules. 

In recent years, studies in the abiotic O$_2$ sources have identified direct O$_2$-production routes from CO$_2$ via intrinsic structural transformation, without requiring atomic O intermediates. Theoretical calculations first predicted the possibility of producing O$_2$ from CO$_2$~\cite{HWANG2000,Grebenshchikov2013}, which was then demonstrated by the experimental work of~\citeauthor{Zhou2014}~\cite{Zhou2014}. 
This study revealed the O$_2$ formation pathway via vacuum ultraviolet photodissociation of CO$_2$ (h$\nu$ + CO$_2$ $\rightarrow$ C + O$_2$). 
Later on, dissociative electron attachment (e$^-$ + CO$_2$ $\rightarrow$ C$^-$ + O$_2$) and He$^+$ ion-induced charge exchange (He$^+$ + CO$_2$ $\rightarrow$ O$_2$ + C$^+$ + He) processes have been identified as further pathways to produce neutral O$_2$ from CO$_2$~\cite{Wang2016,Zhi2023}. Moreover, the dissociation processes of doubly- and singly-charged CO$_2$ ions forming cationic O$_2$$^+$ (CO$_2$$^{2+}$/CO$_2$$^+$ $\rightarrow$ C$^+$/C + O$_2$$^+$) have also been identified, which can be efficiently converted into abiotic O$_2$ through electron neutralization or charge exchange~\cite{Zhao2019,larimian2017,Zhi2023,laksman2012,Monteiro2024,Eland2014,Ganguly2022,Kumar2024,Dong2024}. 
These results significantly advance our understanding of the Earth's atmospheric evolution, and provide more insights into planetary atmospheres and interstellar chemical processes. During the O$_2$/O$_2$$^+$ generation, a fascinating question arises concerning how a linear O-C-O molecule could eliminate its central C atom to produce O-O bonding molecules.

For the excited CO$_2$ molecules, there are two mechanisms for such central C elimination: one involves dissociation from a
transition state (TS) accessed by prior bending deformation of the molecule and the other
invokes breakage of one C-O bond followed
by intramolecular roaming~\cite{Zhou2014,Grebenshchikov2013,HWANG2000}. Since its first observation in 2004, roaming reactions has been established as a generic aspect of chemical reactivity~\cite{Townsend2004}. Evidence of roaming dynamics have been observed across diverse molecular systems involving various fragments~\cite{Mitchell2020,Casey2021,Tomoyuki2020,Wang2021,Wallner2022,Li2024,Chang2023}. In the O roaming pathway, after one C-O bond breaks, the O atom without enough energy to escape and orbits the remaining fragment until encountering a reactive site to forming a C-O-O isomer through intramolecular abstraction. In the case of ionized CO$_2$, only the former mechanism has been proposed~\cite{larimian2017,Kumar2024}, whereas the roaming pathway remains unknown both experimentally and theoretically.
Recent electron-impact absolute cross section measurements indicate that double ionization of CO$_2$ molecule can contribute considerably to the O$_2$$^+$ production~\cite{Monteiro2024}. However, the mechanistic and dynamical details of this O$_2$$^+$ formation process are still need to be elucidated.    

In this work, we present new insights into the production of O$_2$$^+$ from doubly ionized CO$_2$ by electron-impact with energy of 200 eV, which corresponds to the energy measured for free electrons in Mar's atmosphere~\cite{Mitchell2016,Peterson2016}. We also notice here that the doubly-charged ions are readily formed by direct ionization and Auger processes in the planetary atmospheres as that are always exposed to ionizing radiations~\cite{Mitchell2016,Peterson2016,Thissen2011,Feldman1975,stverak2015}. We study the two-body (C$^+$ + O$_2$$^+$) fragmentation channel of CO$_2$ using a multi-particle coincidence momentum spectrometer via a cold target recoil ion momentum spectroscopy or reaction microscope~\cite{DORNER2000,JUllrich2003} combined with a pulsed photoemission electron source, in which the momentum vectors of two fragment ions are measured in coincidence with the scattered electron~\cite{Ren2022,Zhou2025}. The ionization mechanism is determined by measuring the projectile energy loss spectrum in association with the C$^+$ + O$_2$$^+$ channel, which shows two obvious structures in the kinetic energy release (KER) spectrum. Furthermore, we performed potential energy surface (PES) calculations and classic trajectory simulations of the C$^+$ + O$_2$$^+$ channel, which enable us to elucidate the dynamical details of the reaction pathways, particularly for a new roaming mechanism. 

\begin{figure}[h]
\includegraphics[width=0.49\textwidth]{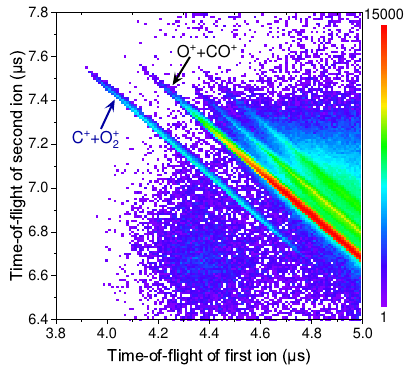}    
\caption{Measured two ions TOF correlation map of the two-body dissociation of CO$_2$$^2$$^+$, in which the  O$^+$ + CO$^+$ and C$^+$ + O$_2$$^+$ channels are clearly identified.
\label{fig1}}
\end{figure}

In the experiment, the time-of-flight (TOF) and position of fragment ions and electrons were recorded with two position- and time-sensitive multihit detectors, which allows the offline determinations of mass-to-charge ratios, momentum vectors, and kinetic energies for all three charged particles (see Supplemental Material). We first characterize O$_2$$^+$ production from CO$_2$$^{2+}$ dication by the ion-ion TOF correlation map. As shown in Fig.~\ref{fig1}, two correlated fragments originating from the same precursor molecule show a clear TOF correlation due to momentum conservation, which manifests as a sharp line with a slope of $-$1. The most intense channel in Fig.~\ref{fig1} is assigned to O$^+$ + CO$^+$, while the adjacent channel on the left corresponds to C$^+$ + O$_2$$^+$, which provides definitive evidence of O$_2$$^+$ production from CO$_2$$^{2+}$. The yield of the C$^+$ + O$_2$$^+$ channel is approximately 0.174$\%$ that of the O$^+$ + CO$^+$ channel, consistent with the previous strong-field laser and ion collision experiments~\cite{larimian2017,Kumar2024}.
	
\begin{figure}[ht]
\includegraphics[width=0.49\textwidth]{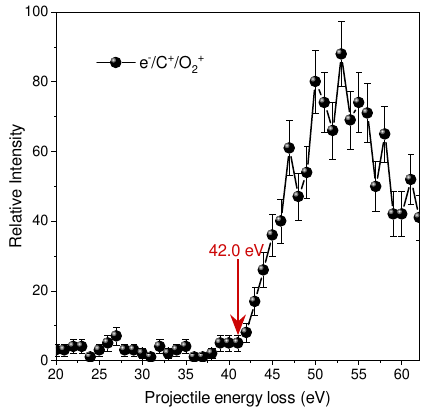}
\caption{Projectile energy loss spectrum of the C$^+$ + O$_2$$^+$ channel obtained by electron-ion-ion triple-coincidence measurement. The red arrow indicates the onset of the spectrum.
\label{fig2}} 
\end{figure}

To identify the initial state of CO$_2$$^{2+}$ dications that produce O$_2$$^+$, we measured the projectile energy loss ($\it{E}$$\rm_{loss}$) spectrum of the C$^+$ + O$_2$$^+$ channel. Here, the $\it{E}$$\rm_{loss}$ is defined as the energy difference between incident electron ($\it{E}$$_0$) and scattered electron ($\it{E}$$_1$), which determines the minimum ionization energy for a specific channel. Figure~\ref{fig2} displays the measured $\it{E}$$\rm_{loss}$ spectrum of the C$^+$ + O$_2$$^+$ channel, which is calibrated with the known double-ionization energy of CO$_2$ ($\it{E}$$\rm_{loss}$[CO$_2$$^{2+}$] = 37.3 eV)~\cite{Eland2003,Zhang2013}. The onset of the $\it{E}$$\rm_{loss}$ spectrum (as indicated by the red arrow at $\sim$ 42.0 eV) corresponds to the $^3$$\Pi$$_u$ dicationic excited state, which is attributed to the ejections of one electron from the highest occupied molecular orbital (HOMO) and the other one from HOMO-2~\cite{Zhang2013,ELAND2016}. In electron-impact experiments, this state can be accessed either through direct double ionization or via inner-shell ionization of the 3$\sigma$$_1$$_g$/2$\sigma$$_2$$_u$ orbitals followed by Auger processes~\cite{Hwang1996}. Recent cross-section analysis of O$_2$$^+$ formed by electron-impact with CO$_2$ indicate that parent CO$_2$$^{2+}$ ions originate predominantly from the Auger processes~\cite{Monteiro2024}.

To form O$_2$$^+$ from CO$_2$$^{2+}$ dication, a bond rearrangement pathway has been proposed in which the symmetrical stretching and bending of the two C-O bonds lead to the reduced $\angle$O-C-O and subsequently new bond formation between the two O atoms~\cite{larimian2017,Kumar2024}. However, the dynamics governing the formation of O$_2$$^+$ via CO$_2$$^{2+}$ remain elusive, particularly for the question of whether the roaming mechanism can play a role in this reaction. This pathway necessitates the cleavage of one C-O bond and the formation of a C-O-O$^{2+}$ intermediate with an intramolecular O abstraction prior to the C$^+$ + O$_2$$^+$ dissociation process. In order to elucidate the dynamics of O$_2$$^+$ formation from CO$_2$$^{2+}$, we have performed {\it ab initio} calculations on the reaction coordinate of CO$_2$$^{2+}$ $\rightarrow$ C$^+$ + O$_2$$^+$. Our calculations were carried out using the Minnesota 2006 hybrid meta-exchange-correlation functional with double the amount of non-local exchange (2X) (M062X) method and the aug-cc-pVTZ basis set~\cite{Becke1993,lee1988}. The relative energies of all TSs and intermediate states (INs) are corrected with zero point energy (ZPE) using the CCSD/aug-cc-pVTZ method~\cite{going2014}. The intrinsic reaction coordinate (IRC) calculations were also performed to verify reaction pathways linking TS to local minima (see Supplemental Material). 

\begin{figure}[ht]
\includegraphics[width=0.45\textwidth]{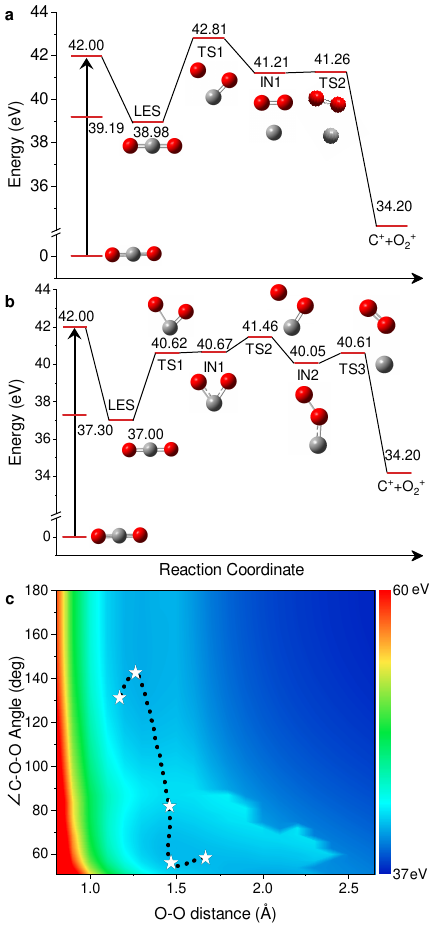}
\caption{Calculated reaction pathways with the selected TSs and INs along the potential energy curves of the C$^+$ + O$_2$$^+$ channel from the singlet (a) and triplet (b) states of CO$_2$$^2$$^+$. (c) Two-dimensional PES as a function of the $\angle$C-O-O angle and O-O distance. The dotted line indicates the roaming path, in which the white five-pointed stars represent the TSs and INs shown in (b).
\label{fig3}}
\end{figure}

The calculated potential energies for the reaction from the CO$_2$$^{2+}$ dicationic states to the final products C$^+$ + O$_2$$^+$ are presented in Fig.~\ref{fig3}. For the reaction corresponding to the singlet state (a$^1$$\Delta$$_g$$^+$) of CO$_2$$^{2+}$ (shown in Fig.~\ref{fig3}a), we first determine the vertical ionization energy to be 39.19 eV, which is in good agreement with the previous studies~\cite{Eland2003,Zhang2013}. After double ionization to the singlet state, the CO$_2$$^{2+}$ dicationic molecule can relax to the local lowest energy state (LES), then rearrange significantly the molecular structure with stretching and bending deformations through a transition state TS1, during which the distance between the two O atoms decreases to 1.68 $\rm\AA$. In the intermediate state IN1, the CO$_2$$^{2+}$ dication adopts a triangular structure, with the $\angle$O-C-O angle and O-O bond distance being 41.8$^{\circ}$ and 1.18 $\rm\AA$, respectively. This structural rearrangement leads to the formation of a new O-O covalent bond. The bending deformation has been proven to be very fast and can be completed in just a few femtoseconds~\cite{laksman2012}. The intermediate state IN1 is characterized by the weakened C-O bonds, which consequently leads to fragmentation with an extremely low energy barrier of only 0.05 eV (TS2), resulting in the final products C$^+$ + O$_2$$^+$. This small energy barrier suggests that the fragmentation process is highly favorable and proceeds efficiently once the IN1 state is reached. 

In addition to the bond rearrangement pathway revealed in Fig.~\ref{fig3}a, we identify a further roaming mechanism for formation of O$_2$$^+$ from CO$_2$$^{2+}$. As shown in Fig.~\ref{fig3}b, the roaming pathway is initiated by population on the triplet state (X$^3$$\Sigma$$_g$$^-$) of CO$_2$$^{2+}$. The calculated vertical ionization energy of X$^3$$\Sigma$$_g$$^-$ state amounts to 37.3 eV, which is consistent with the double ionization threshold of CO$_2$ determined by the experimental and theoretical studies~\cite{Eland2003,Zhang2013}. Upon double ionization and relaxation to the local LES, the CO$_2$$^{2+}$ dication then crosses the first transition state (TS1) and reaches to a metastable intermediate state IN1, which is characterized by a remarkable structural rearrangement. At this stage, the initially linear O-C-O backbone experiences a drastic angular deformation, bending from 180$^{\circ}$ to 67.5$^{\circ}$, along with a substantial contraction of the O$\cdot\cdot\cdot$O internuclear distance from its equilibrium value 2.32 $\rm\AA$ to 1.44 $\rm\AA$. The CO$_2$$^{2+}$ subsequently progresses toward the second transition state (TS2). Here, the charge redistribution results in a significant weakening of the C-O bond, ultimately leading to its complete cleavage. This quasifree O atom, not having enough energy to escape, rotates around the C-O moiety and ultimately recombines with it at a suitable reaction site (IN2), forming a C-O-O$^{2+}$ isomer. This newly formed O-O bond adopts an equilibrium distance of 1.25 $\rm\AA$, and the accompanying $\angle$C-O-O angle opens to 143.5$^{\circ}$. The C-O-O$^{2+}$ dication further crosses an energy barrier of approximately 0.6 eV (TS3), leading to the C-O bond breaking and the formation of C$^+$ + O$_2$$^+$ ion pairs. 
It is noted in Fig.~\ref{fig3}b that the potential energies of TSs and INs from TS1 to TS3 are very close, thereby posing a flat potential energy surface (PES), which is essential for O roaming and ultimately abstracting another O to form oxygen molecule. 
Here, all TSs lie energetically below the initial $^3$$\Pi$$_u$ state, enabling a barrierless roaming process. The electronically excited CO$_2$$^{2+}$ ($^3$$\Pi$$_u$) undergoes internal conversion to the triplet state PES, the O atom then wanders around varied configuration spaces of the flat potential energy regions and forms a C-O-O$^{2+}$ intermediate prior to the final products C$^+$ + O$_2$$^+$. 

\begin{figure}[ht]
\includegraphics[width=0.49\textwidth]{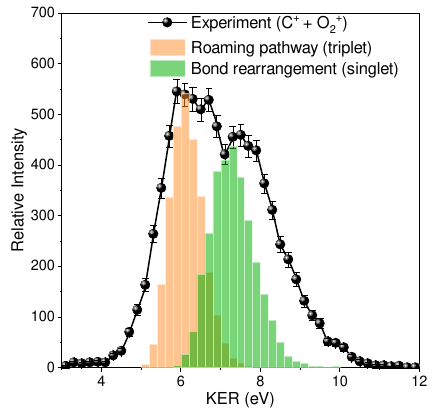}
\caption{Measured KER spectrum for the C$^+$ + O$_2^+$ channel. The orange and green bar distributions represent the KER spectra calculated with the Coulomb explosion model applied to charge separation using the triplet-state (TS3) and singlet-state (TS2) transition structures, respectively.
\label{fig4}}
\end{figure}

The roaming pathway is further visualized by our calculated two-dimensional PES as a function of the $\angle$C-O-O angle and O-O distance shown in Fig.~\ref{fig3}c. Since the roaming originates from TS1 in Fig~\ref{fig3}b, we map the PES starting from 50$^{\circ}$ of the $\angle$C-O-O angle. The corresponding TSs and INs from Fig.~\ref{fig3}b are also marked in Fig.~\ref{fig3}c to trace the roaming pathway, which exhibits a flat potential energy profile. Along the indicated pathway, potential barriers exist on both sides as the O-O distance increases or decreases. The potential energy of the CO$_2$$^{2+}$ dication decreases as the $\angle$C-O-O angle increases. This leads to the breaking of a C-O bond and subsequent roaming of the O atom around the C-O moiety while keeping the O-O distance constant when increasing the $\angle$C-O-O angle. When the $\angle$C-O-O angle reaches approximately 143$^{\circ}$, the roaming O atom recombines with the C-O moiety, forming a C-O-O structure. Finally, the C-O bond breaks, yielding the final products C$^+$ + O$_2$$^+$. Upon crossing the potential barriers along the O-O coordinate, the C-O-O$^{2+}$ intermediate will rapidly dissociate into O$^+$ + CO$^+$.
		 
						
According to the energy differences between the final transition state and ion products in Fig.~\ref{fig3}, it should lead to different KER distributions for O$_2$$^+$ production from the bond rearrangement and roaming pathways. As shown in Fig.~\ref{fig4}, the different KER distributions are resolved in our measured KER spectrum of the C$^+$ + O$_2$$^+$ channel, which exhibits two dominant peaks centered at about 6.4 eV and 7.6 eV, respectively. We utilize Gaussian fitting to resolve the two KER distributions, obtaining a branching ratio of 55$\%$ : 45$\%$ between the two peaks. The low-energy KER peak agrees well with the result measured by strong-field laser ionization~\cite{larimian2017} and ion collision~\cite{Kumar2024} experiments. This double-peak structure was previously observed in O 1s$\rightarrow$$\pi$$^*$ excited C$^+$ + O$_2$$^+$~\cite{laksman2012}, which confirms that the O$_2$$^+$ production from CO$_2$$^{2+}$ primarily arises from the Auger process in this work. 
The energy difference between TS3 and the ionic products in Fig.~\ref{fig3}b is determined to be 6.41 eV, which is in good agreement with the low-energy KER peak. This indicates that the low-energy KER peak primarily arises from the roaming pathway. While the peak located at KER $\sim$ 7.6 eV is close to the energy difference between TS2 and the ionic products (7.06 eV) in Fig.~\ref{fig3}a, which indicates that the high-energy KER peak is mainly attributed to the bond rearrangement pathway of CO$_2$$^{2+}$. 
		
To further quantify the KER distributions, we carried out molecular dynamics calculations using a classical trajectory method (see Supplemental Material), in which the known molecular structures are used as input parameters to calculate the KER distributions of the ion pair~\cite{Hsieh1995, Hsieh_1997,Eland1994}. During the structural evolution of the singlet and triplet states in CO$_2$$^{2+}$, we obtained two different pre-dissociation TS configurations. For the triplet state, a nearly linear transition state (TS3) was identified, characterized by an associated O-O bond length of 1.15 $\rm\AA$ and a connected C-O bond length of 1.84 $\rm\AA$. The singlet state adopts a bent transition state (TS2), featuring an $\angle$O-C-O bond angle of 93$^{\circ}$ and an O-O bond length of 1.16 $\rm\AA$. By setting the vibrational full width at half maximum (FWHM) of the C-O and O-O bond lengths to 0.3 $\rm\AA$ and the angular FWHM of the $\angle$C-O-O (and $\angle$O-C-O) to 30$^{\circ}$, we calculate the KER distributions for two reaction pathways, which are displayed by the orange (triplet state) and green (singlet state) bar charts in Fig.~\ref{fig4}. The calculated KER distributions response well to the double-peak structure in the experimental spectrum. This result supports again that the two peaks in the KER spectrum are originating from the bond rearrangement and roaming pathways, respectively.

In conclusion, we have studied experimentally and theoretically the production of O$_2$$^+$ from doubly ionized CO$_2$ molecule upon electron-impact, where O$_2$$^+$ can be converted to O$_2$ by electron recombination or charge exchange processes~\cite{Wallner2022}. Using fragment ions and electron coincidence momentum spectroscopy, we have obtained the projectile energy loss spectrum in association with the C$^+$ + O$_2$$^+$ dissociation channel, which provide unambiguous evidence for the mechanism of the $^3$$\Pi$$_u$ excited state-controlled fragmentation of CO$_2$$^{2+}$ dication. Further potential energy and trajectory calculations have elucidated the fragmentation dynamics of the production of O$_2$$^+$ from CO$_2$$^{2+}$, which show good agreement with the measured KER spectrum of C$^+$ + O$_2$$^+$ products. Our studies have revealed the dynamical details of the bond rearrangement pathway, in which an intermediate triangular structure is formed after population of the singlet state of CO$_2$$^{2+}$. This pathway likely contributes to the high energy (7.6 eV) peak of the KER spectrum.

Moreover, we demonstrate a further roaming mechanism for formation of O$_2$$^+$ from CO$_2$$^{2+}$, which is found to be a barrierless pathway. This roaming pathway is initiated by double ionization and relaxation to the triplet state, then a frustrated C-O bond cleavage leaves the O atom without sufficient energy to escape. Subsequently, the O atom wanders around varied configuration spaces of the flat potential energy regions and forms a C-O-O$^{2+}$ intermediate prior to the final products C$^+$ + O$_2$$^+$ (KER $\sim$ 6.4 eV). The abundance of free electrons in interstellar space and the upper atmosphere suggests that the processes revealed here are significant and should be incorporated into atmospheric chemistry models. The present studies provide new insights into the abiotic oxygen generation in planetary atmospheres rich in CO$_2$ and facilitate the identification of hitherto unrecognized pathways of interstellar chemistry and molecular evolution within the universe.

\section*{Acknowledgments}
The work is jointly supported by the National Natural Science Foundation of China under Grants No. 12325406, No. 92261201, No. 12404305, No. 11974272, No. 12120101005, and No. 12175174; the Shaanxi Province Natural Science Fundamental Research Project under Grants No. 2023JC-XJ-03 and No. 23JSQ013; and the China Postdoctoral Science Foundation under Grant No. BX20240286 and No. 2024M7625.

\bibliography{CO2}

\end{document}